# Evaluation of GlassNet for physics-informed machine learning of glass stability and glass-forming ability

Sarah I. Allec[a], Xiaonan Lu[b], Daniel R. Cassar[c,d], Xuan T. Nguyen[e], Vinay I. Hegde[a], Thiruvillamalai Mahadevan[e], Miroslava Peterson[b], Jincheng Du[e], Brian J. Riley[b], John D. Vienna[b], James E. Saal[a],*

[a]*Citrine Informatics, Redwood City, CA 94063, USA*

[b]*Pacific Northwest National Laboratory, Richland, WA 99352, USA* [c]*Ilum School of Science, Brazilian Center for Research in Energy and Materials (CNPEM), 13083-970, Campinas, Sao Paulo, Brazil* [d]*National Institute of Science and Technology on Materials Informatics, Campinas, Brazil* [e]*University of North Texas, Denton, TX 76203, USA*

*Corresponding author. E-mail: jsaal@citrine.io

## Abstract

Glassy materials form the basis of many modern applications, including nuclear waste immobilization, touch-screen displays, and optical fibers, and also hold great potential for future medical and environmental applications. However, their structural complexity and large composition space make design and optimization challenging for certain applications. Of particular importance for glass processing and design is an estimate of a given composition's *glass-forming ability* (GFA). However, there remain many open questions regarding the underlying physical mechanisms of glass formation, especially in oxide glasses. It is apparent that a proxy for GFA would be highly useful in glass processing and design, but identifying such a surrogate property has proven itself to be difficult. While glass stability (GS) parameters have historically been used as a GFA surrogate, recent research has demonstrated that most of these parameters are not accurate predictors of the GFA of oxide glasses. Here, we explore the application of an open-source pre-trained NN model, GlassNet, that can predict the characteristic temperatures necessary to compute GS with reasonable performance and assess the feasibility of using these physics-informed ML (PIML)-predicted GS parameters to estimate GFA. In doing so, we track the uncertainties at each step of the computation – from the original ML prediction errors, to the compounding of errors during GS estimation, and finally to the final estimation of GFA. While GlassNet exhibits reasonable accuracy on all individual properties, we observe a large compounding of error in the combination of these individual predictions for the PIML prediction of GS, finding that random forest models offer similar accuracy to GlassNet. We also breakdown the performance of GlassNet on different glass families and find that the error in GS prediction is correlated with the error in crystallization peak temperature prediction. Lastly, we utilize this finding to assess the relationship between top-performing GS parameters and GFA for two ternary glass systems: sodium borosilicate and sodium iron phosphate glasses. We conclude that to obtain true ML predictive capability of GFA, significantly more data needs to be collected.

**Introduction**

Glassy materials form the basis of many modern applications, including nuclear waste immobilization, touch-screen displays, and optical fibers,[1] and also hold great potential for future medical and environmental applications.[2] However, the structural complexity and unfathomably large composition space of these materials make design and optimization challenging for certain applications: Glass compositions are not restricted to stoichiometric rules, meaning that their compositional ranges are continuous. This is both a blessing and a curse, as the opportunities for unique properties are virtually limitless, yet efficiently searching such a huge compositional space for materials that have relatively ill-defined structure-property relationships (with respect to crystalline materials) is extremely challenging.

Of particular importance for glass processing and design is an estimate of a given composition's *glass-forming ability* (GFA). In theory, any material can vitrify if it can be cooled from the molten state to its glass transition temperature ($T_g$) quickly enough to prevent crystallization. The ease with which one can form a glass for a given material composition, *i.e.*, how slow a material can be cooled without significant crystallization, is what is meant by the term GFA. It is determined by the material's *critical cooling rate* ($R_c$), practically defined to be the minimum cooling rate necessary to form a glass piece with a crystalline volume fraction ($X_c$) less than some threshold (usually taken to be in the range of $10^{-2}$ to $10^{-6}$).[3] In practice, a material's GFA determines how easily one will be able to make a glass in the lab, as materials with low GFA often require specific rapid cooling methods such as twin roller quenching. There remain many open questions regarding the underlying physical mechanisms of nucleation, crystal growth, and glass formation, especially in multi-component oxide glasses.[1] Furthermore, the determination of $R_c$ is quite tedious, requiring either the time-consuming construction of time-temperature-transformation (TTT) diagrams or other methods[4-5] that were shown to exhibit high uncertainties by Jiusti *et al.*[6] Therefore, accurate prediction of GFA for unknown compositions is currently not possible due to a lack of physical understanding of the mechanisms at play and a dearth of robust $R_c$ data.

It is apparent that a proxy for GFA would be highly useful in glass processing and design, but identifying such a surrogate property has proven itself to be difficult, particularly for oxide glasses. While glass stability (GS) parameters, which measure the resistance of a glass to crystallization upon reheating, have historically been used as a GFA surrogate, recent research has demonstrated that most of these parameters, of which there are a few dozen in the literature,[7-31] are not accurate predictors of the GFA of oxide glasses.[3, 32-36] Furthermore, while GS determination is less time-consuming than GFA determination, it requires characteristic glass temperatures (*e.g.,* glass transition temperature, $T_g$, and crystallization peak temperature, $T_c$), thus necessitating the production of a glass sample. To date, the largest study evaluating the relationship between GS parameters and GFA for oxide glasses contained twelve glasses: six silicate, five borate, and a germania glass.[3] Out of the thirty five GS parameters evaluated, the authors identified nine GS parameters with a correlation mode (based on coefficients of determination) greater than 0.7 for predicting GFA. The fact that the largest study on this topic consisted of such a small dataset is reflective of the difficulties in obtaining high quality data for both GFA and GS.

In such a data-limited field, the application of machine learning to inform experiments and accelerate materials design comes with its own set of unique challenges. The large neural network (NN) models underlying modern deep learning techniques are data-hungry, requiring on the order of (at least) thousands of data points. Furthermore, the accuracy of machine learning (ML) models generally depends on the quality of the training data,[37] where quality is affected by experimental uncertainty and error, relevance of features to the target property, and diversity of the dataset. For glasses, experimental uncertainty can be large for the characteristic temperatures underlying GS and GFA, as the value of any of these properties not only depends on composition but also experimental conditions, processing, and equipment. In addition to these variations in target property values, identification of input features that are predictive of the target properties is difficult, particularly when it comes to the reduced number of structural features available for glasses as opposed to crystalline materials. Lastly, the small amount of data available for GFA severely limits the diversity of training data available to the ML model, thus limiting the generality of the model.

While there currently are no datasets sufficient for ML model training and prediction of GFA directly, here we explore the application of an open-source pre-trained NN model, GlassNet,[38] that can predict the characteristic temperatures necessary to compute GS with reasonable performance, and assess the feasibility of using these physics-informed ML (PIML)-predicted GS parameters to estimate GFA. In doing so, we track the uncertainties at each step of the computation – from the original ML prediction errors, to the compounding of errors during GS estimation, and finally to the final estimation of GFA. While GS parameters can be computed directly from the GlassNet-predicted characteristic temperatures, GlassNet does not contain values for the parameters needed to compute GFA for any glass system other than sodium borosilicates. Instead, we utilize the 12-glass dataset used by Jiusti et al.[3] (hereafter referred to as the Jiusti dataset) and predictions on the two ternary systems to assess the errors in the last step (GFA estimation). We have also assessed the feasibility of using the newly proposed liquid-property-based *Jezica* parameter, $JEZ = \frac{\eta(T_l)}{T_l^2}$,[6] where $\eta(T_l)$ is the shear viscosity at the liquidus temperature $T_l$, for estimating GFA. The benefit of using *Jezica* over GS parameters as a GFA surrogate property is that it only requires properties of the liquid, *i.e.,* a glass piece does not need to be experimentally produced in order to measure its fundamental properties and compute it. Furthermore, Jiusti et al.[3] demonstrated that *Jezica* has a higher predictive capability for GFA than the majority of GS parameters for their 12-glass dataset.

While GlassNet exhibits reasonable accuracy on all individual properties used in the various GS definitions, we observe a large compounding of error in the combination of these individual predictions for the PIML-prediction of GS, finding that random forest (RF) models offer similar accuracy to GlassNet. We also breakdown the performance of GlassNet on different glass families (*e.g.,* silicates, borates, phosphates) and find that the error in GS prediction is correlated with the error in crystallization peak temperature ($T_c$) prediction. Lastly, we utilize this finding to assess the relationship between the top-performing GS parameters and GFA for a well-studied glass system, sodium borosilicates, as well as sodium iron phosphate glasses, a glass system of interest for its application in nuclear waste disposal. Across all the test sets considered in this work (GlassNet test set, Jiusti dataset, and the two ternary systems), the most quantitatively reliable GS parameter predicted by ML and also predictive of GFA across glass families is $H'(T_c)$, which is solely dependent on $T_g$ and $T_c$. However, when attempting to identify the glass-forming region of a specific glass system, it is unclear which parameter can accurately predict the changes in GFA as a function of composition, as we observe that ML-predictions of *Jezica* can correctly identify the glass forming region of the sodium borosilicate ternary[42], while ML-predictions of $H'(T_c)$ correctly identify the glass forming region of the sodium iron phosphate ternary[43]. It is unclear whether this difference arises from the compositional representation of these two families in the GlassNet training data or from differences in the physical glass formation mechanisms of these two families. Lastly, we re-emphasize that the results here show that the accuracy of PIML GS predictions is dependent on the ML-predicted value of $T_c$, implying that gathering more $T_c$ data could help improve ML GS predictions. Nonetheless, to offer true ML predictive capability of GFA across different glass systems (*e.g.,* to compare phosphates to silicates) and within a single glass system (*e.g.,* identifying the glass-forming region in a ternary phase diagram), significantly more data needs to be collected that contains all characteristic temperatures for a given glass. Because the results here demonstrate that random forests, from which robust uncertainty estimates are much more easily attained than neural networks, are sufficient for the prediction of GS- and GFA-relevant properties, we suggest that an uncertainty-based surrogate modeling approach may be useful in guiding the data generation and collection process.

## Methods

*ML models*

The GlassNet model is capable of predicting 85 glass properties from composition alone, all of which are described in the original reference.[38] The model is trained on the SciGlass database, where duplicate compositions have been aggregated based on their median property values (see original reference[38] for further data cleaning details). In the versions used in this work, GlassNet 0.4.6 and 0.5.0, there is a multi-task version of GlassNet, as well as single-task neural networks for properties that are predicted better by single-task neural networks (STNN), and the use of these STNNs can be toggled by setting an input parameter to the GlassNet class. In multi-task learning, the training process is shared across all tasks, opening up the possibility of learning relationships between targets because the loss is taking into account the error on all tasks simultaneously. On the other hand, in single-task learning, there is a single training process for each task, so such joint learning is not possible. A schematic that visualizes these two types of learning is shown in **Figure 1**. We also compare the performance of GlassNet to single-task random forest (RF) models, as RFs have been very successful on several materials datasets, particularly small materials datasets. All RF models are trained on subsets of the GlassNet training data (*e.g.,* for a given property, say $T_g$, the RF model was trained on the subset of GlassNet training data containing values for $T_g$ (*i.e.,* rows without a measured value for $T_g$ were dropped)). Similarly, all RF test sets are subsets of the GlassNet test data. GS and GFA are then computed from these values and are taken to be the "true" GS and GFA values according to the definitions in **Table 2**; predicted GS and GFA values come from the combination of the predicted values. The input features for all RF models are the same features used for the GlassNet model. The RF models were built with scikit-learn[39] version 1.2.0, using the default parameters. For all results except the predictions on the ternary systems, we used

GlassNet version 0.4.6, as this model was trained on the GlassNet training data only (*i.e.,* the dataset used to perform model evaluation in the original GlassNet paper[38]). This allowed us to make fair comparisons between our own RF models and GlassNet by using GlassNet's exact train-test split (*i.e.,* all accuracy metrics and parity plots are *from predictions on the GlassNet test* set using models *trained on the GlassNet training set*). In the most recent version of GlassNet, version 0.5.0, the model was trained on the entire cleaned SciGlass database (*i.e.,* the GlassNet training *and* test data combined) - this production model was used to predict on the ternary systems in **Figures 9-11**.

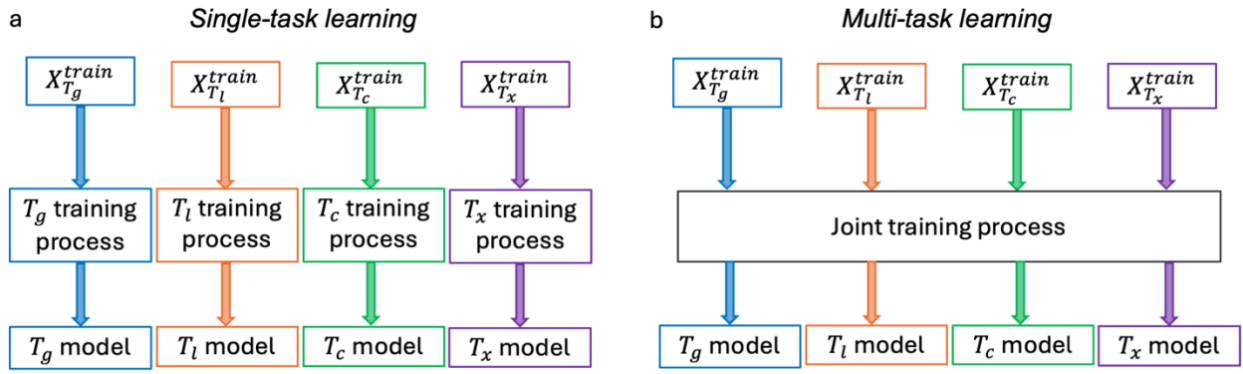

**Figure 1**. Schematic of (a) single-task learning and of (b) multi-task learning.

*Assessment of Jezica parameter*

The *Jezica* parameter was developed to be a liquid-based predictor, as opposed to the glass-based GS parameters, of GFA, depending only on the shear viscosity of the liquid, $\eta$, at its liquidus temperature, $T_l$.[6] Because the *Jezica* parameter relies on $\eta(T_l)$, a property that SciGlass does not contain, we computed $\eta(T_l)$ by regressing the MYEGA (Mauro–Yue–Ellison–Gupta–Allan) equation[40] on of $\eta$ at various temperatures, which SciGlass does contain:

$$\log_{10} \eta(T) = \log_{10}(\eta_\infty) + (12 - \log_{10}(\eta_\infty))\frac{T_{12}}{T} \exp\left[\left(\frac{m}{12-\log_{10}(\eta_\infty)} - 1\right)\left(\frac{T_{12}}{T} - 1\right)\right], \quad (1)$$

In the construction of this dataset, we only selected liquids for which the measured $T_l$ is between the lowest and highest temperature at which viscosity had been measured.

*GFA estimation*

Throughout this paper, we define GFA as the inverse log of the *critical cooling rate* $R_c$:

$$GFA = \frac{1}{\log_{10} R_C}, \qquad (2)$$

The critical cooling rate is the minimum cooling rate necessary to form a glass piece with a crystalline fraction ($X_s$) less than some threshold (usually taken to be in the range of $10^{-2}$ to $10^{-6}$). In simpler terms, $R_c$ is the slowest a liquid can be cooled without significant crystallization. In a similar vein to Jiusti *et al.*,[3] we take the definition of $R_c$ for heterogeneous nucleation, which is typically dominant in oxide glasses over homogeneous nucleation:

$$R_c = \frac{T_l - T_{\max (U)}}{t_n} \qquad (3)$$

$$t_n = \sqrt{\frac{X_s}{\pi N_s U_{max}^2}}, \qquad (4)$$

where $T_l$ is the liquidus temperature, $U_{max}$ is the maximum crystal growth velocity, $T_{\max (U)}$ is the temperature of maximum crystal growth velocity, $N_s$ is the number of sites per unit area inducing heterogeneous crystallization, and $X_s$ is the maximum allowed crystallized fraction. Here, we have used the same order of magnitude estimates as Jiusti *et al.*[3] for $X_s$ and $N_s$, taking $X_s = 10^{-2}$ and $N_s = 10^3$.

Because GlassNet does not contain values for the parameters needed to compute GFA for any glass system other than sodium borosilicates, we utilize the 12-glass dataset from Table 2 of Jiusti *et al.*[3], hereafter referred to as the Jiusti dataset. This dataset contains all the parameters needed to compute the various GS parameters, *Jezica*, and GFA for 12 glasses (six silicates, five borates and the germania glass). We have verified that this dataset is well within the property distributions of the GlassNet training data for all properties except $\log(U_{max})$, as demonstrated by its property distributions as well as GlassNet's prediction accuracy on these glasses (see **Figures S1-S2**).

*CALPHAD calculations*

We performed CALPHAD calculations of ternary silicate systems using FToxid, the FACT oxide database with the FACTSage v8.2 software[41]. Liquidus temperatures of a wide range of compositions in these systems were obtained based on the calculated phase diagrams.

# Results

We first assess the accuracy of GlassNet and RF for predicting all properties needed for computing GS and GFA. All GS parameters are computed from a combination of some subset of the following characteristic temperatures: glass transition temperature $T_g$, liquidus temperature $T_l$, crystallization peak temperature $T_c$, and crystallization onset temperature $T_x$. Here, we only compute the GS parameters identified by Jiusti *et al.*[3] as predictive of GFA with a linear correlation $r^2$ mode (the most frequent $r^2$ for bootstrapped linear regressions) of at least 0.7; these parameters and their definitions are displayed in column 1 of **Table 2**. According to equations 2-4, GFA depends on $T_l$, maximum crystal growth velocity $U_{max}$, and the temperature of maximum crystal growth velocity $T_{\max(U)}$. All temperatures in this work are in Kelvin, $U_{max}$ is in m/s, and viscosities are in Pa·s. In **Figure 2**, we show parity plots for all of these properties predicted by the full multi-task GlassNet ("No STNN or RF"), by GlassNet's STNNs ("With STNN"), and by the RF models ("With RF"). We quickly observe that *i)* the temperature models (**Figure 2a-d**) are very accurate (see **Table 1** for coefficient of determination, $R^2$, values), *ii)* the accuracy of the RF models for all properties are comparable to those of GlassNet with STNNs (see **Table 1** for further verification), and *iii)* the amount of data containing values for all of the GFA-relevant properties

($T_l$, $U_{max}$, $T_{\max(U)}$; **Figure 2e-f**) is very much reduced relative to the entire GlassNet dataset. We also note that we have plotted the average variation in each property ($\sigma_{SciGlass}$) for a given composition in the SciGlass database as the shaded green region around the parity line, as many compositions are duplicated in SciGlass. Here, we have taken the 75$^{th}$ percentile of the standard deviation for each property as this variance, computed over all duplicated compositions. We have also tabulated these values in **Table 1**, observing that all temperatures have $\sigma_{SciGlass} \leq 10$ K, except for $T_c$ and $T_x$.

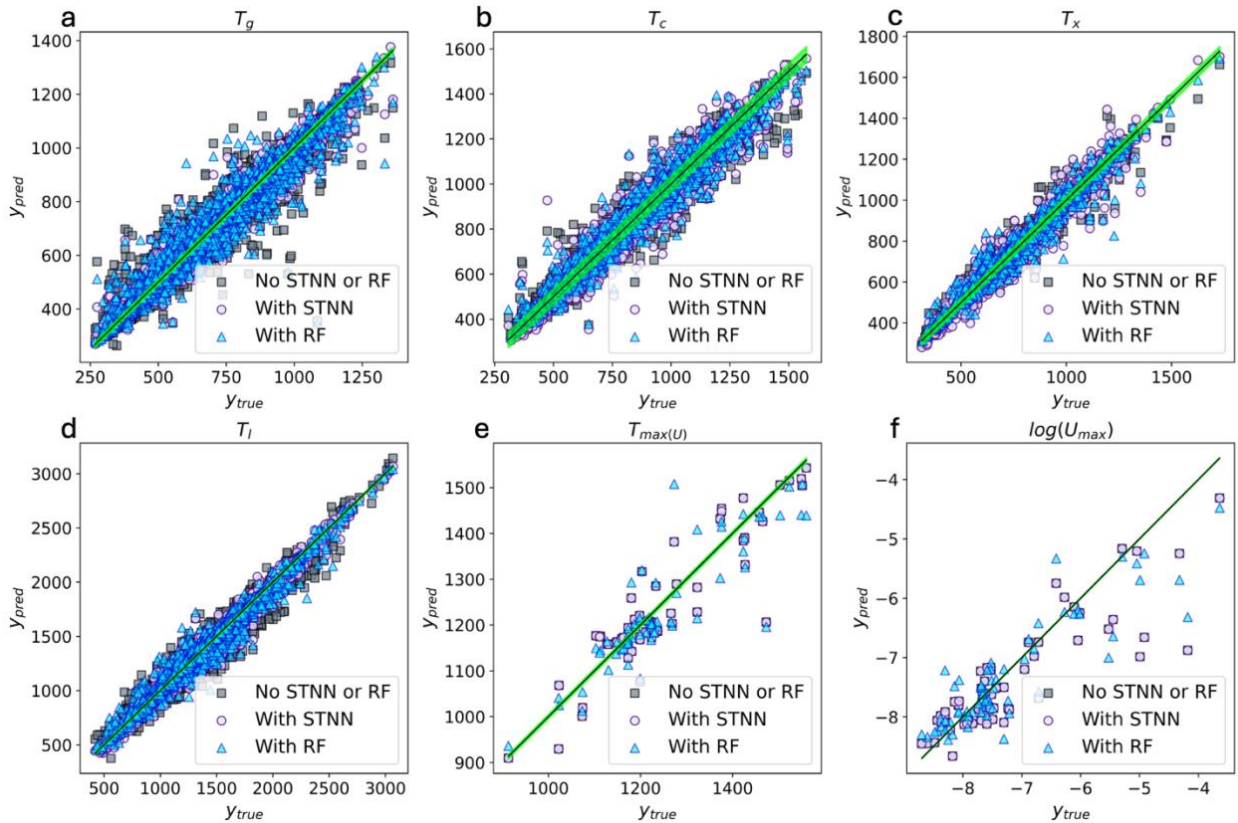

**Figure 2.** Parity plots ($y_{pred}$ vs $y_{true}$) for the various properties needed to compute (a-d) GS and (e-f) GFA, as predicted by GlassNet and RF models. Predictions by multi-task GlassNet, STNN GlassNet, and RF are represented by gray squares, lavender circles, and cyan triangles, respectively. Green shaded region denotes the standard deviation in these temperatures for a given composition in the SciGlass database, where we have taken the 75$^{th}$ percentile of the standard deviations over all duplicated compositions. Note: $T_{\max(U)}$ and $\log(U_{max})$ do not have

STNN models in GlassNet, so the markers for multi-task GlassNet and STNN GlassNet overlap. All temperatures are in Kelvin and $U_{max}$ is in m/s.

**Table 1.** Variations in each property for duplicated compositions in SciGlass ($\sigma_{SciGlass}$) and coefficients of determination ($R^2$) for each model. All temperatures are in Kelvin and $U_{max}$ is in m/s. Note: $T_{\max(U)}$ and $\log(U_{max})$ do not have STNN models in GlassNet, so they do not have values for $R^2_{STNN}$. The variation in RF $R^2$ values was less than or equal to 0.001 for all models, computed over ten different model trainings.

| Property | $\sigma_{SciGlass}$ | $R^2_{No\_STNN}$ | $R^2_{STNN}$ | $R^2_{RF}$ |
|---|---|---|---|---|
| $T_g$ | 10 K | 0.95 | 0.98 | 0.97 |
| $T_c$ | 33 K | 0.93 | 0.95 | 0.96 |
| $T_x$ | 21 K | 0.94 | 0.94 | 0.96 |
| $T_l$ | 8 K | 0.94 | 0.97 | 0.97 |
| $T_{\max(U)}$ | 4 K | 0.81 | N/A | 0.76 |
| $\log(U_{max})$ | 9E-10 | 0.72 | N/A | 0.77 |

Using these predictions, we have computed the GS parameters of **Table 2** as well as GFA, with **Table 2** containing the coefficients of determination for each model (columns 2-4), as well as their predictive capability for GFA, $r^2_{GFA}$ (column 5; values taken from Jiusti *et al.*[3] Table 3). Across the board, there is a significant drop in accuracy for the GS parameters, with the highest $R^2$ (0.63) achieved using GlassNet with STNNs for predicting $\gamma(T_c)$. In the case of GFA, GlassNet and RF give $R^2$ values of 0.68 and 0.77, respectively, both of which are comparable to the $R^2$ values of GlassNet and RF for $\log(U_{max})$ (0.72 and 0.77, respectively). For the rest of this work, we take the top three ML-predicted GS parameters in terms of PIML prediction accuracy and predictive capability for GFA – $K_w(T_c)$, $\gamma(T_c)$, and $H'(T_c)$ – for further analysis. The parity plots for these three GS parameters along with GFA are shown in **Figure 3**. Based on **Table 2** and **Figure 3**, the most promising models are RF for predicting $H'(T_c)$ and GlassNet STNNs or RF for predicting $\gamma(T_c)$. We note that *while the GFA model has the highest R² values, the training dataset is very small for ML (205 data points) and thus the model should not be expected to be generalizable*. However, the compounding of errors for this model is significantly lower than that of any GS models, as the GFA $R^2$ values are comparable to the $R^2$ values of the underlying properties. This is very likely due to the fact that the GlassNet data subsets for GFA-relevant properties - $T_l$, $U_{max}$, $T_{\max(U)}$ – contain similar glasses, while the subsets for GS-relevant properties are varied, as will be discussed next.

Table 2. Definitions of GS parameters, coefficients of determination ($R^2$) for each model for GS parameters and GFA, and bootstrapped correlation mode for GS predictive capability of GFA[3]. The top three ML-predicted GS parameters in terms of PIML prediction accuracy and predictive capability for GFA – $K_w(T_c)$, $\gamma(T_c)$, and $H'(T_c)$ – are in green text. Note: We give a standard deviation for $K_H(T_c)$'s $R^2_{RF}$ because it is the only GS parameter that exhibits significant variance in $R^2$. The variation in RF $R^2$ values was less than or equal to 0.05 for all other models, computed over ten different model trainings.

| GS parameter | $R^2_{No\_STNN}$ | $R^2_{STNN}$ | $R^2_{RF}$ | $r^2_{GFA}$ |
|---|---|---|---|---|
| $K_w(T_c) = \dfrac{T_c - T_g}{T_l}$ | 0.31 | 0.44 | 0.45 | 0.81 |
| $\gamma(T_c) = \dfrac{T_c}{T_g + T_l}$ | 0.41 | 0.63 | 0.59 | 0.78 |
| $K_w(T_x) = \dfrac{T_x - T_g}{T_l}$ | 0.17 | 0.11 | 0.02 | 0.77 |
| $H'(T_x) = \dfrac{T_x - T_g}{T_g}$ | 0.22 | 0.20 | 0.35 | 0.77 |
| $K_H(T_x) = \dfrac{T_x - T_g}{T_l - T_x}$ | 0.05 | 0.07 | 0.05 | 0.75 |
| $K_H(T_c) = \dfrac{T_c - T_g}{T_l - T_c}$ | 0.32 | 0.18 | 0.14 (±0.28) | 0.74 |
| $H'(T_c) = \dfrac{T_c - T_g}{T_g}$ | 0.28 | 0.45 | 0.60 | 0.71 |
| $\Delta T_{rg} = \dfrac{T_x - T_g}{T_l - T_g}$ | 0.24 | 0.26 | 0.25 | 0.70 |
| $K_{CR} = \dfrac{T_l - T_x}{T_l - T_g}$ | 0.24 | 0.26 | 0.25 | 0.70 |
| $GFA$ | 0.65 | 0.68 | 0.77 | 1 |

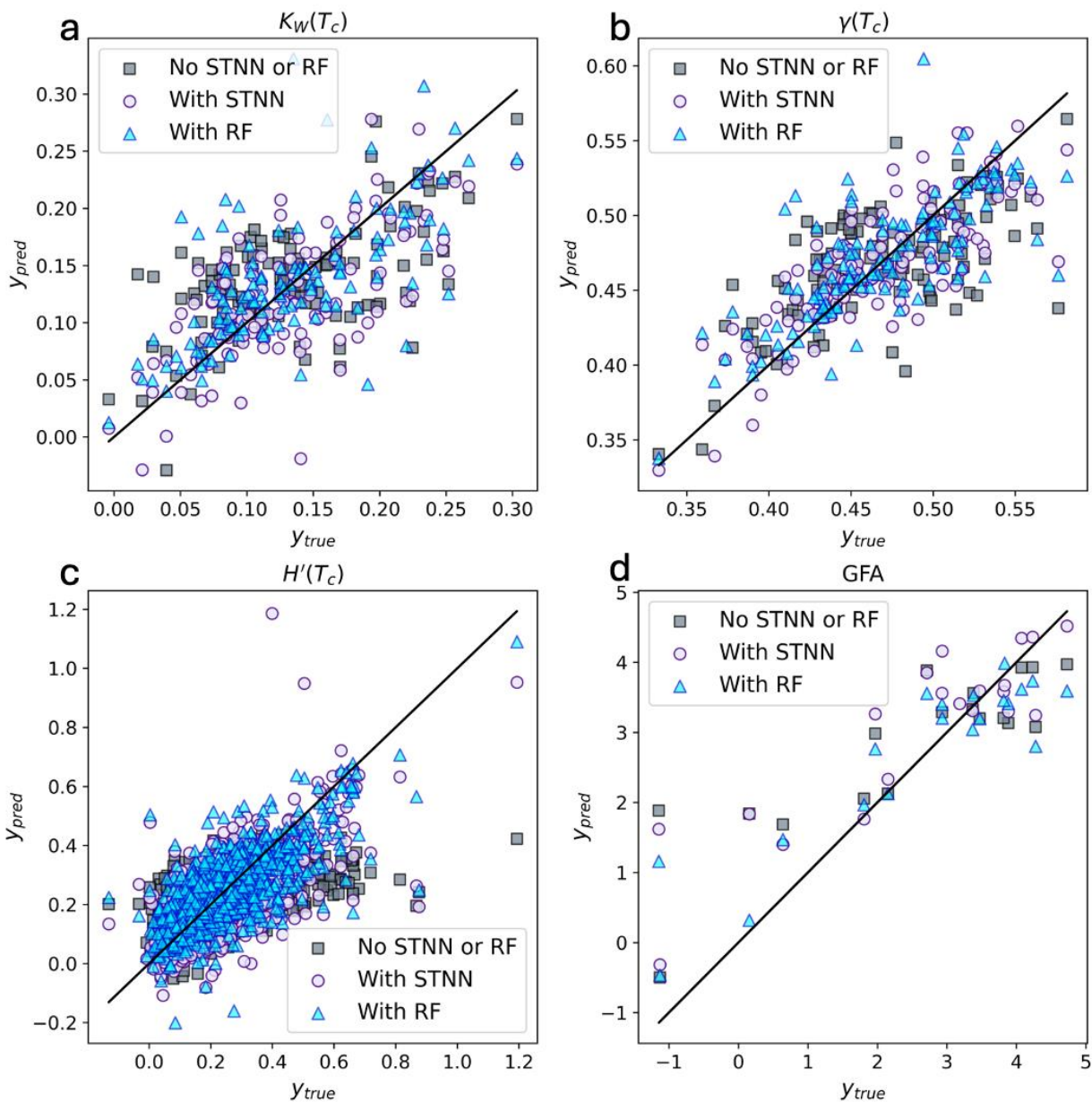

**Figure 3.** Parity plots ($y_{pred}$ vs $y_{true}$) for (a-c) the top three ML-predicted GS parameters and (d) GFA. Predictions by multi-task GlassNet, STNN GlassNet, and RF are represented by gray squares, lavender circles, and cyan triangles, respectively.

To understand the observed drop in performance when combining ML predictions that have high individual accuracies (**Table 1**), we examine the types of glasses contained in and the size of each property subset. In **Figure 4a**, we plot the most common elements (defined by the number of glasses containing at least 0.1 of a given element in atomic fraction) and their average atomic fractions, for each property subset relevant for GS. As expected, we see that the network former elements are represented the most in all characteristic temperature subsets, with the $T_c$ and $T_x$ subsets containing a wider diversity of network former elements (Ge, Te, F, Se in addition to Si, P, and B). Na is one of the pre-dominantly represented network modifiers, as demonstrated by its presence in this plot but also by its representation in **Figure 5d-f**. Of more importance for ML prediction accuracy is the fact that the $T_c$ and $T_x$ subsets are relatively small, as shown in **Figure 4b** – 9% and 5% of the total GlassNet training set, respectively. As a result, the intersection of the $T_c$ or $T_x$ subset with one or two of the other subsets is further reduced, meaning that each temperature model is trained on different glasses. Hence, the combination of each these models results in a drop in accuracy, as each model is learning the relationship between composition and GS for different glass types.

However, this is not the case for GFA, as shown by similar plots **Figure 4c-d**, where each of the properties pre-dominantly contains glasses with Si and Na as the former and modifier elements, respectively. Thus, *while the training datasets for the GFA-relevant properties are generally smaller than those for the GS-relevant properties, the GFA predictions are more accurate in terms of their capability to predict on similar glasses. However, the models for $log(U_{max})$ and $T_{max\,(U)}$ should not be assumed to give similar prediction accuracy on other types of glasses.* In fact, if we count how many glasses contain Si and Na as the *dominant* former and modifier (*i.e.,* glasses for which the atomic fraction for Si and Na is greater than the summed fractions for all other formers and modifiers in the glass, respectively), we observe that the subset containing all GFA-relevant properties is dominated by these glasses, as shown in **Figure 5c,f**. Thus, GlassNet and RF models for predicting $T_l$, $\log(U_{max})$, and $T_{\max\,(U)}$ on this data may be reliable for sodium silicates, is most likely even less reliable for lithium, potassium, and calcium silicates (see bottom right plot of **Figure 5f**), and most likely completely unreliable for any other type of glass.

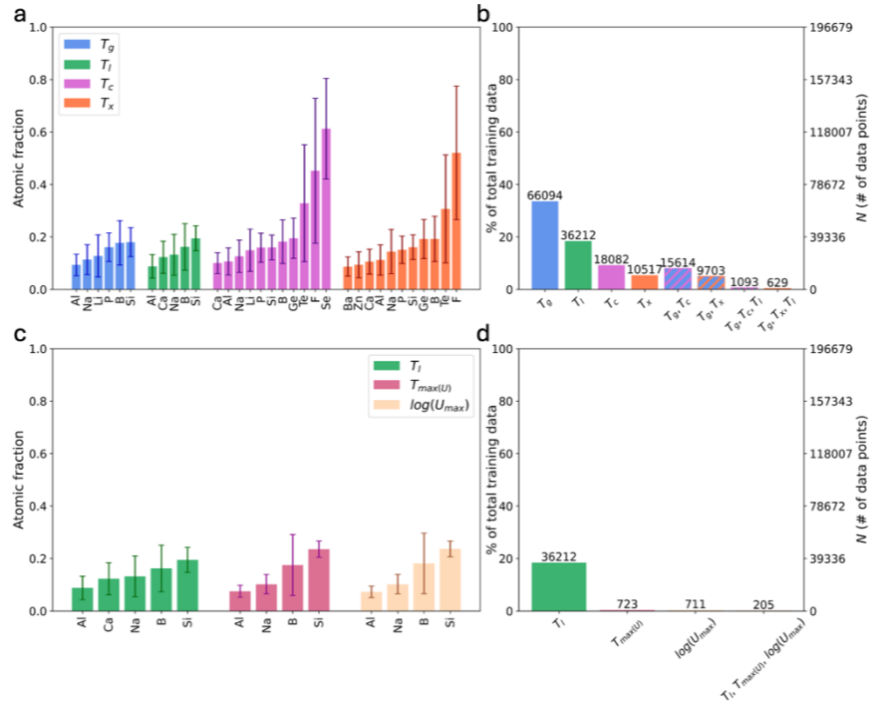

**Figure 4.** (a) Atomic fractions of commonly represented elements in and (b) size of each GS-relevant property subset; (c) atomic fractions of commonly represented elements in and (d) size of each GFA-relevant property subset. Error bars represent the standard deviations of the properties in each subset. Subsets with multiple properties (*e.g.,* $T_g, T_c$) are defined as the glasses in the GlassNet training data with values for all properties (*i.e.,* the intersection of the subsets for each individual property). The scales for *y* were chosen to ensure a fair comparison of atomic fractions between (a) and (c), and to show the extent of the dataset size reduction from the entire dataset ("100% of total training data") in each subset for (b) and (d).

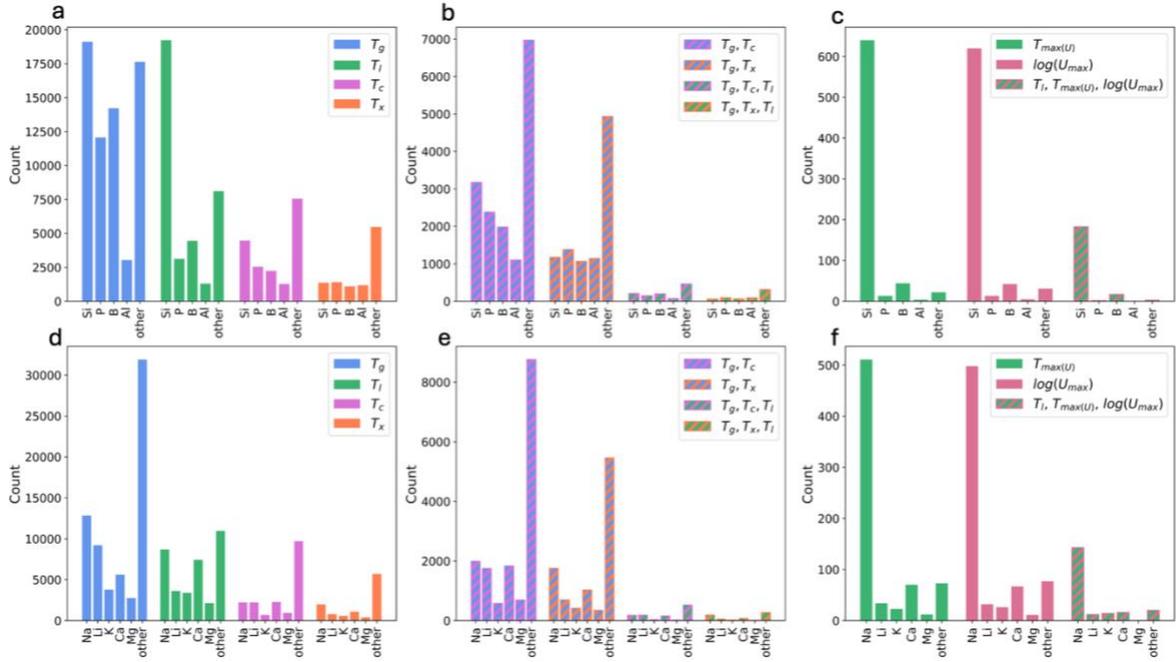

**Figure 5.** Counts of glasses with a *dominant* former or modifier (*i.e.,* glasses where the atomic fraction of a single former/modifier is greater than the sum of all other formers/modifiers): (a), (d) Former, modifier counts for GS-relevant individual property subsets; (b), (e) former, modifier counts for GS-relevant combined property subsets; (c), (f) former, modifier counts for GFA-relevant individual property (excluding $T_l$) and combined property subsets. Glasses without a dominant former or modifier are grouped under "other".

According to **Figures 4-5**, the GlassNet training data for the characteristic temperatures is much more diverse than the data for the GFA-relevant properties. To get a sense of how well GlassNet performs on predicting GS for different glass families, **Figure 6** contains parity plots for the characteristic temperatures, colored by family, with the corresponding $R^2$ values given in **Table 3**. **Figure 6** and **Table 3** are for predictions from GlassNet STNNs; a similar figure and table for RF are in **Figure S3** and **Table S1**. "Other" in **Figure 6** and **Table 3** refer to glasses where the dominant network former element either is not one of Si, P, B, or Al, or where there is no single dominant network former element. For all temperatures, the GlassNet STNNs perform the best on the aluminates subset of the test set, and this performance is not related to their representation in the training data, as aluminates are the smallest subset for all temperatures except $T_x$. We also see that the total $R^2$ over all glasses (those in the last row of **Table 3** and column 2 of **Table 1**) seem to reflect GlassNet's performance on the aluminates and "other" glasses, as the other glass families have significantly lower $R^2$ values for all temperatures except $T_g$.

We now see how these differences in temperature prediction accuracy are reflected in the GS parameter accuracy, which is shown as glass-family-specific parity plots in **Figure 7a-c** and in their mean absolute error ($MAE$) and $R^2$ values in **Table 4**; a similar figure and table for RF are in **Figure S4** and **Table S2**. Because of the reduction in subset size for glasses containing $T_g$, $T_c$, and $T_l$, it is difficult to draw conclusions regarding the GS prediction accuracy by family, except that $K_W(T_c)$ and $\gamma(T_c)$ seem reliable for silicates (and most likely across different network modifiers as well, as shown in **Figure 4**). This could be due to their being a significant representation of silicates in each of these temperature subsets, as shown in **Figure 5** and **Table 3**. To see how the errors in the individual temperature predictions affect the errors in the GS computations, we computed the correlation (Kendall's rank correlation coefficient) between the residuals for each temperature and the residuals for each GS parameter, where the residual is given by $y_{pred} - y_{true}$. The only temperature whose residuals are correlated with the GS residuals is $T_c$, with correlations in the range $0.6 - 0.78$ for certain glass families. The correlations between the $T_c$ residuals and the GS residuals are clearly seen in the plots in **Figure 7d-f,** where the GS residuals are plotted against the $T_c$ residuals. This could be because the $T_c$ models are the least accurate of the three temperatures used to compute these GS parameters (**Table 3**).

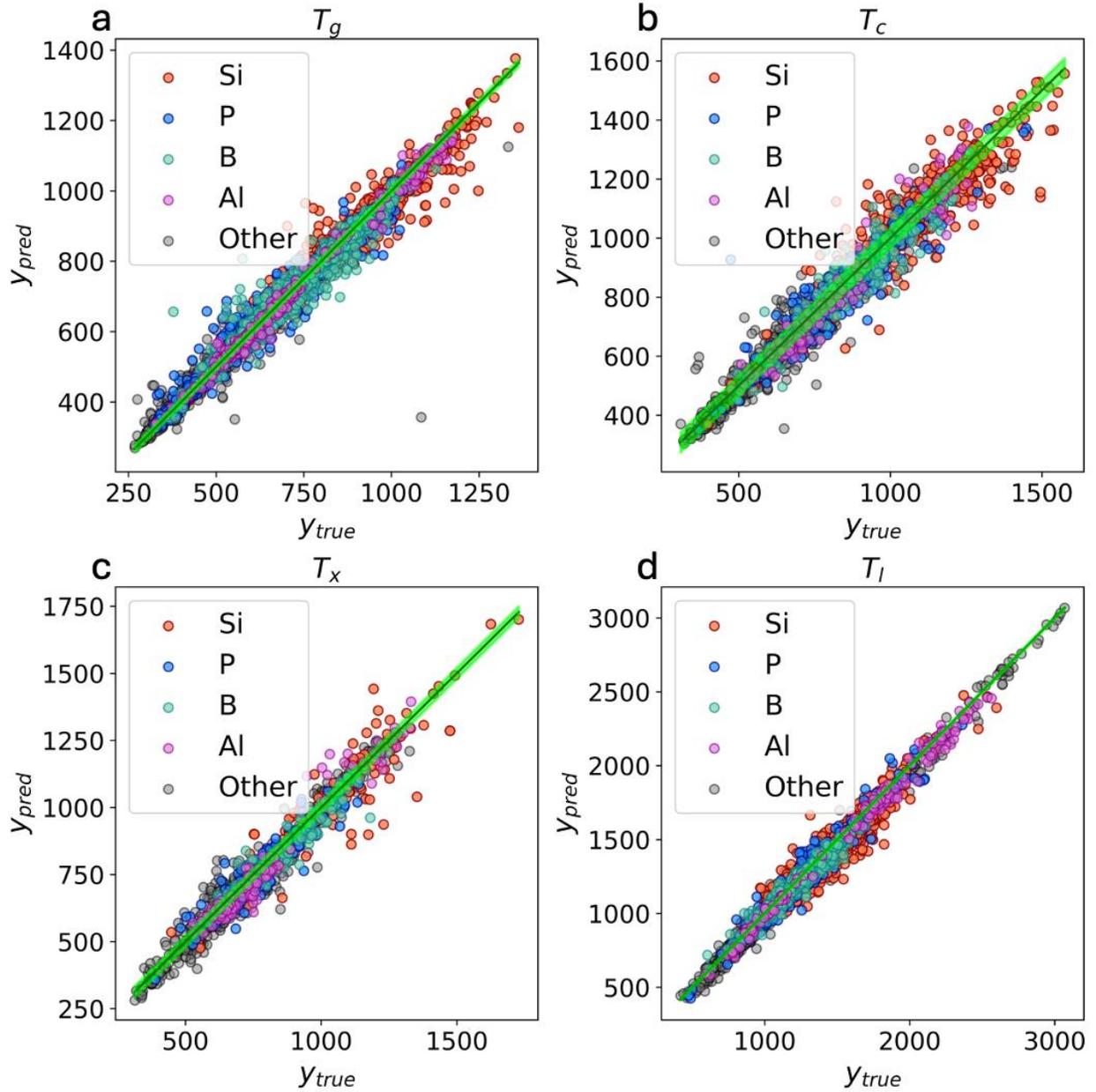

**Figure 6.** Parity plots ($y_{pred}$ vs $y_{true}$) for the GS-relevant properties (characteristic temperatures) broken down by family, as predicted by GlassNet. All temperatures are in Kelvin and $U_{max}$ is in m/s. Green shaded region denotes the standard deviation in these temperatures for a given composition in the SciGlass database, where we have taken the 75[th] percentile of the standard deviations over all duplicated compositions.

**Table 3.** Number of training data points, $N_{T_i}$, and coefficients of determination, $R^2_{T_i}$, for each characteristic temperature, broken down by glass family. The last row contains the corresponding values for all glasses for reference.

| Former element | $N_{T_g}$ | $R^2_{T_g}$ | $N_{T_c}$ | $R^2_{T_c}$ | $N_{T_x}$ | $R^2_{T_x}$ | $N_{T_l}$ | $R^2_{T_l}$ |
|---|---|---|---|---|---|---|---|---|
| **Si** | 19,118 | 0.96 | 4,471 | 0.85 | 1,364 | 0.83 | 19,238 | 0.90 |
| **P** | 12,074 | 0.96 | 2,538 | 0.87 | 1,402 | 0.87 | 3,117 | 0.95 |
| **B** | 14,228 | 0.95 | 2,235 | 0.86 | 1,100 | 0.88 | 4,449 | 0.93 |
| **Al** | 3,036 | 0.99 | 1,274 | 0.94 | 1,174 | 0.95 | 1,300 | 0.99 |
| **Other** | 17,638 | 0.97 | 7,564 | 0.93 | 5,477 | 0.93 | 8,108 | 0.99 |
| *All* | 66,094 | 0.98 | 18,082 | 0.95 | 10,517 | 0.94 | 36,212 | 0.97 |

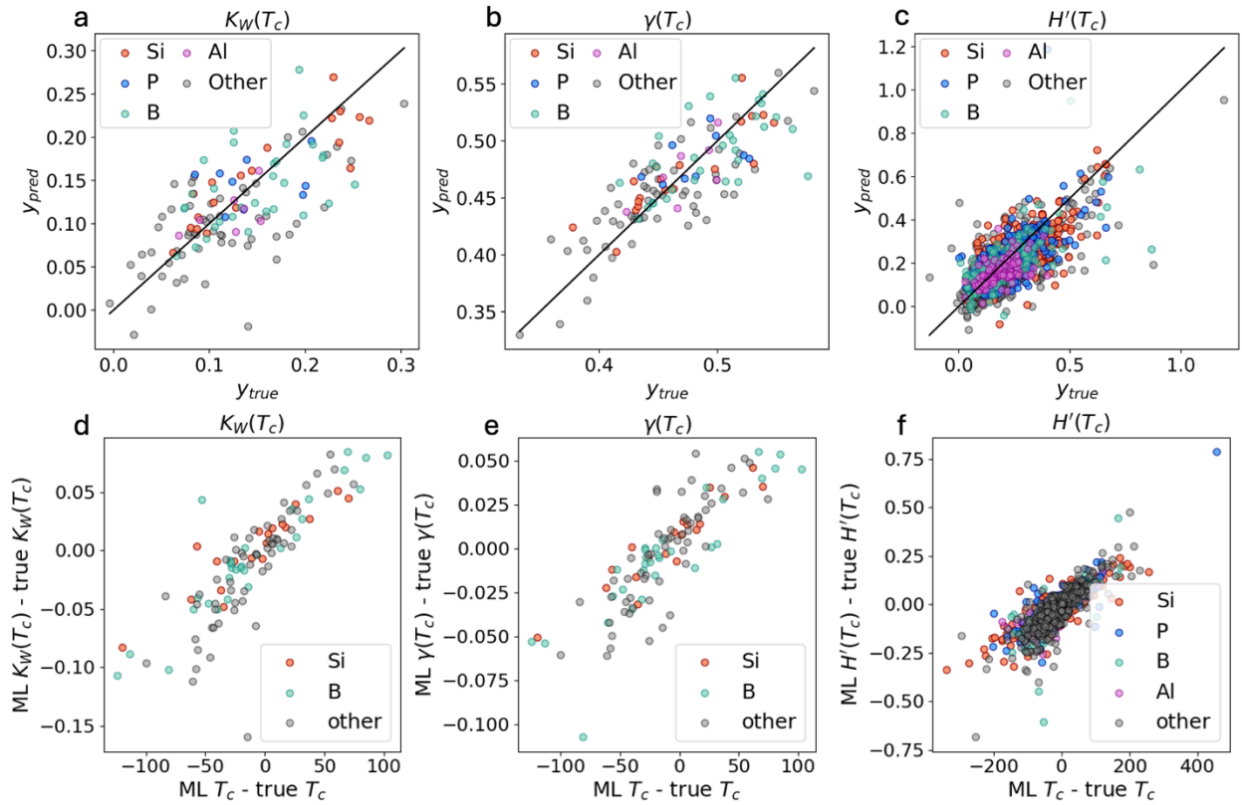

**Figure 7.** (a-c) Parity plots ($y_{pred}$ vs $y_{true}$) for the top three ML-predicted GS parameters broken down by family, as predicted by GlassNet. (d-f) GS residuals vs. $T_c$ residuals for (d) $K_w(T_c)$, (e) $\gamma(T_c)$, and (f) $H'(T_c)$, demonstrating the correlation between the two.

**Table 4.** Coefficients of determination, $R^2_i$, and mean absolute errors, $MAE_i$, for each GS parameter, broken down by glass family. Because $R^2$ values are not well-defined for small

dataset sizes, no $R^2$ are given for GS parameters with a test set size of less than ten. The last row contains the corresponding values for all glasses for reference.

| Former element | $R^2_{K_w}$ | $MAE_{K_w}$ | $R^2_\gamma$ | $MAE_\gamma$ | $R^2_{H'}$ | $MAE_{H'}$ |
|---|---|---|---|---|---|---|
| Si | 0.78 | 0.026 | 0.75 | 0.019 | 0.44 | 0.067 |
| P | NA | 0.038 | NA | 0.022 | 0.33 | 0.061 |
| B | 0.04 | 0.039 | 0.31 | 0.025 | 0.47 | 0.061 |
| Al | NA | 0.021 | NA | 0.017 | 0.1 | 0.048 |
| Other | 0.38 | 0.038 | 0.65 | 0.025 | 0.44 | 0.065 |
| All | 0.44 | 0.036 | 0.63 | 0.024 | 0.45 | 0.063 |

We now evaluate the accuracy of GlassNet for predicting *Jezica*, which utilizes $T_l$ predictions and predictions on the viscosity at various temperatures, which are fed into the MYEGA regression to obtain $\eta(T_l)$. Thus, there are two layers of PIML: one in the computation of $\eta(T_l)$ and one in the computation of *Jezica*. **Figure 8** contains the parity plots for each of these predictions, where we see that the accuracy of *Jezica* prediction is primarily determined by the accuracy of $\eta(T_l)$ prediction. We expect that the MYEGA regression would have a compounding of errors, which gives rise to the low accuracy of this prediction. We also note that there is a significant difference in the prediction accuracy of silicates and that of any other glass family with a single dominant former. In fact, GlassNet does not seem to have any predictive capability on the phosphates and borates in this dataset, as exhibited by the flatness of those families' parity lines. However, while the silicates in **Figure 8** are predicted well, we have also compared GlassNet $T_l$ predictions on several silicate ternary systems to $T_l$ values computed from CALPHAD, as shown in **Figure 9**, and see that the $T_l$ prediction accuracy of GlassNet can vary significantly within silicates, with particularly poor performance for MgO-Al$_2$O$_3$-SiO$_2$ across all $T_l$ and for CaO-FeO-SiO$_2$ in the lower $T_l$ range.

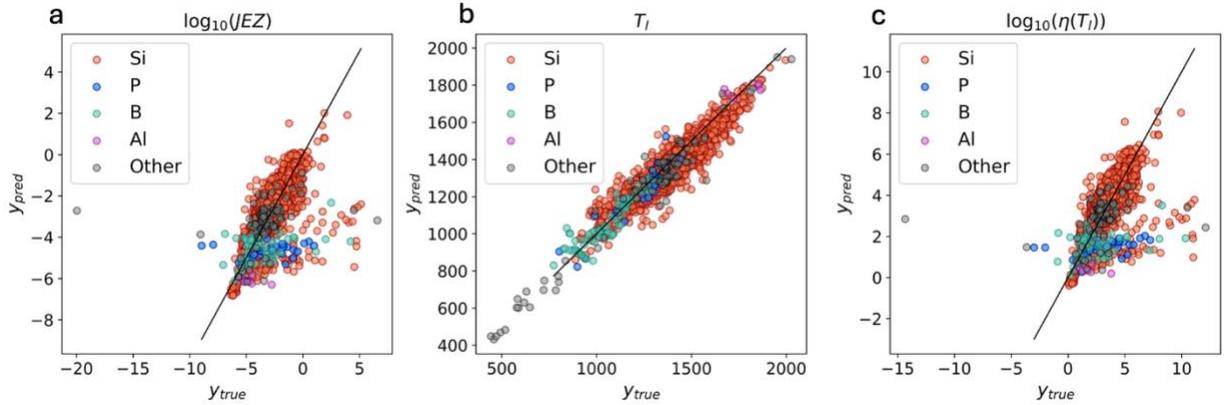

**Figure 8.** Parity plots ($y_{pred}$ vs $y_{true}$) for the (a) *Jezica* parameter, (b) $T_l$, and (c) $\eta(T_l)$, broken down by family, as predicted by GlassNet. ) $T_l$ is in Kelvin and $\eta(T_l)$ is in Pa·s.

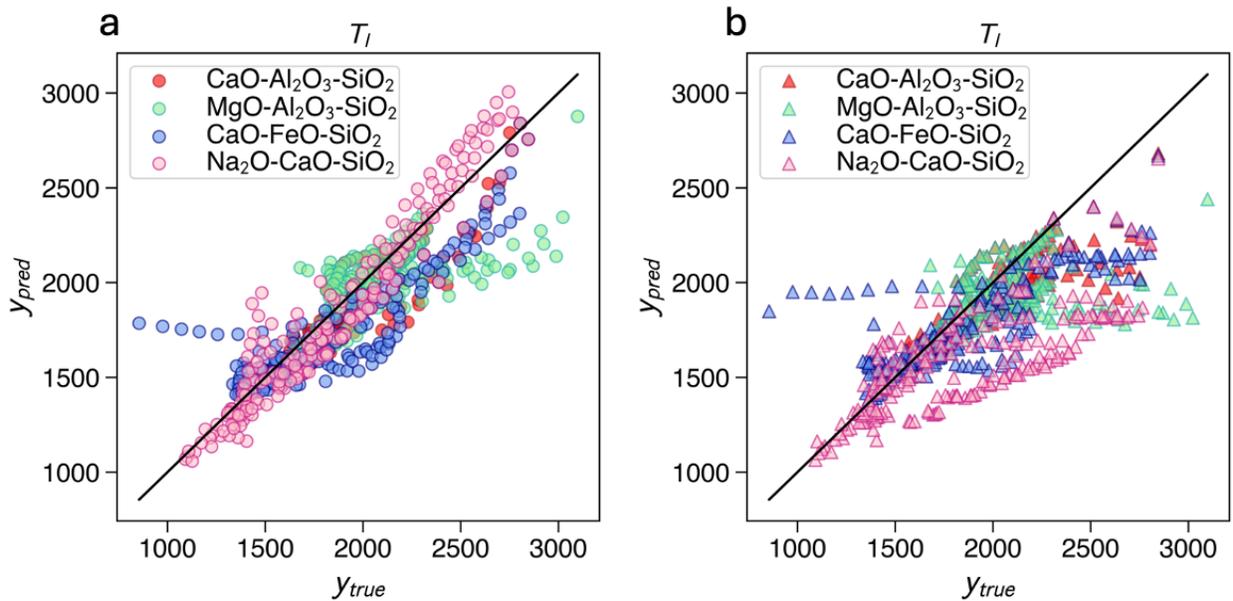

**Figure 9.** Parity plots ($y_{pred}$ vs $y_{true}$) for $T_l$ predicted by GlassNet for four silicate ternary systems, where the true values were calculated with CALPHAD. $T_l$ is in Kelvin.

We now turn to the viability of GlassNet-predicted GS parameters and the *Jezica* parameter for identifying glass-forming regions in two ternary systems: sodium borosilicates and sodium iron phosphates. To evaluate the viability of GlassNet and RF for PIML predictions of *Jezica*, we have used GlassNet and RF predictions for $T_l$ and GlassNet predictions for $\eta(T_l)$ from the MYEGA regression described in the Methods section. Thus, there is some mixing of GlassNet and RF predictions in what we refer to as the RF models for *Jezica*. While Jiusti et al.[3] demonstrated a correlation between *Jezica* and GFA for the Jiusti dataset, it was unclear to us whether or not PIML-predicted *Jezica* values are also correlated with GFA. In **Figure S5**, we demonstrate that the PIML-predicted *Jezica* values are indeed correlated with their true GFA values for this glass dataset. However, this correlation should be dependent on the accuracy of the *Jezica* predictions, and **Figures 8-9** show that this accuracy varies significantly across glass systems. To see the implications of this, in **Figures 9** and **10** we have plotted ternary phase diagrams for $Na_2O$-$B_2O_3$-$SiO_2$ and $Na_2O$-$Fe_2O_3$-$P_2O_5$, respectively, as predicted by GlassNet STNNs and RF models, where we have also included RF models for *direct* prediction of GS, *i.e.,* instead of training individual models for $T_g$, $T_l$, and $T_c$ and combining these predictions to compute the GS parameters (*indirect* GS prediction), we took the GlassNet training subset containing all of the relevant temperatures for a given GS parameter and trained an individual model for that GS parameter – further details are provided in the SI. These *direct* models are not an apples-to-apples comparison with the other models but are included because they perform well on $Na_2O$-$Fe_2O_3$-$P_2O_5$. The true phase diagram for each system is shaded in gray and bounded by a black line. In line with **Figure 9**, we see that the *Jezica* parameter best predicts the glass-forming region of $Na_2O$-$B_2O_3$-$SiO_2$, while $H'(T_c)$ best predicts the glass-forming region of $Na_2O$-$Fe_2O_3$-$P_2O_5$, while all other parameters perform relatively poorly for both systems. It is unclear whether or not these results are due to the compositional breakdown of the relevant training data for these two glass systems, as silicates are better represented than phosphates for all properties. Even if we look at the number of glasses in each property subset containing Na, B, Si and Na, Fe, P, as shown in **Figure 11**, there is more data in each property for Na$_2$O-B$_2$O$_3$-SiO$_2$. It is also very likely that if the *Jezica* prediction for phosphates was as accurate as it is for silicates, that *Jezica* would also identify the glass-forming region of Na$_2$O-Fe$_2$O$_3$-P$_2$O$_5$. Another possibility is that this difference in "most predictive GFA parameter" is due to some difference in the glass formation mechanism for these two families of glasses, but investigation of this possibility is outside the scope of this study. Regardless, these ambiguous results highlight the need for more data containing *all* GS-relevant properties for each glass.

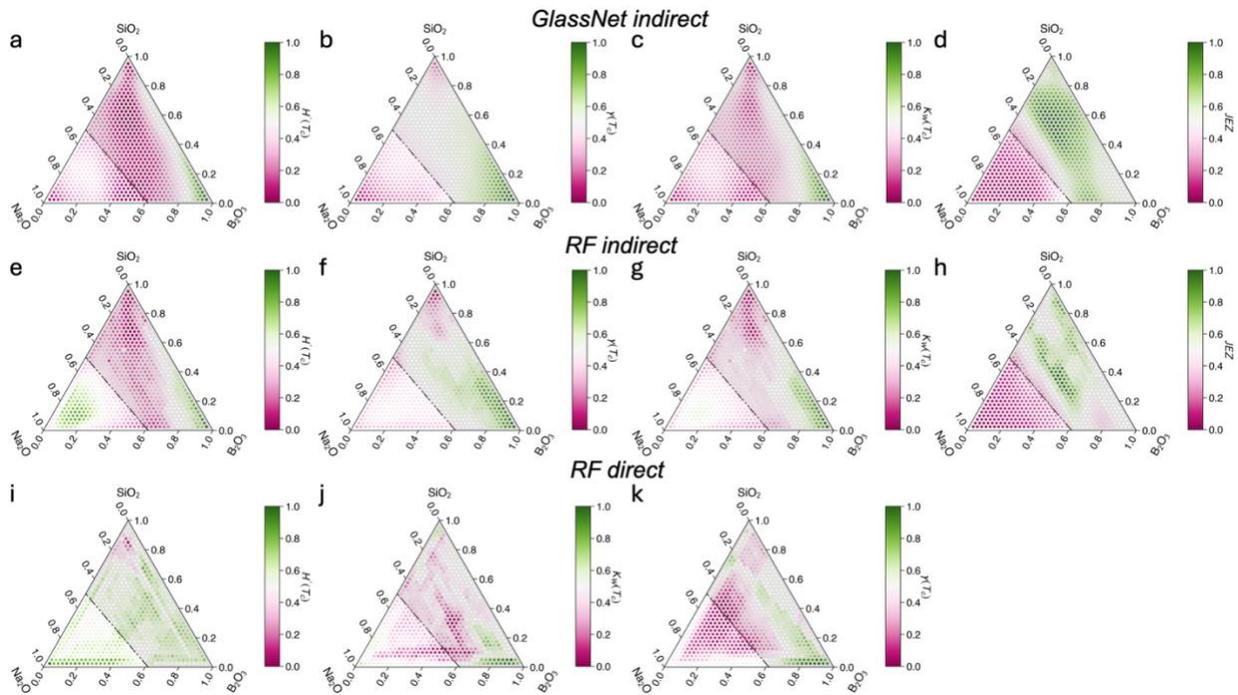

**Figure 10.** PIML predictions of GS parameters and *Jezica*, projected onto the Na$_2$O-B$_2$O$_3$-SiO$_2$ ternary: (a-d) GlassNet (indirect) predictions, (e-h) RF indirect predictions, and (i-k) RF direct predictions. The true glass-forming region[42] is shaded in gray and bounded by a black line. Indirect predictions here refer to predictions made from a combination of the predictions for the characteristic temperatures and log($U_{max}$); direct predictions refer to models that directly predict

a GS parameter. Note that there is no direct prediction of *Jezica* because we do not have $\eta(T_l)$ values in the training data and cannot compute "true" *Jezica* parameters for training.

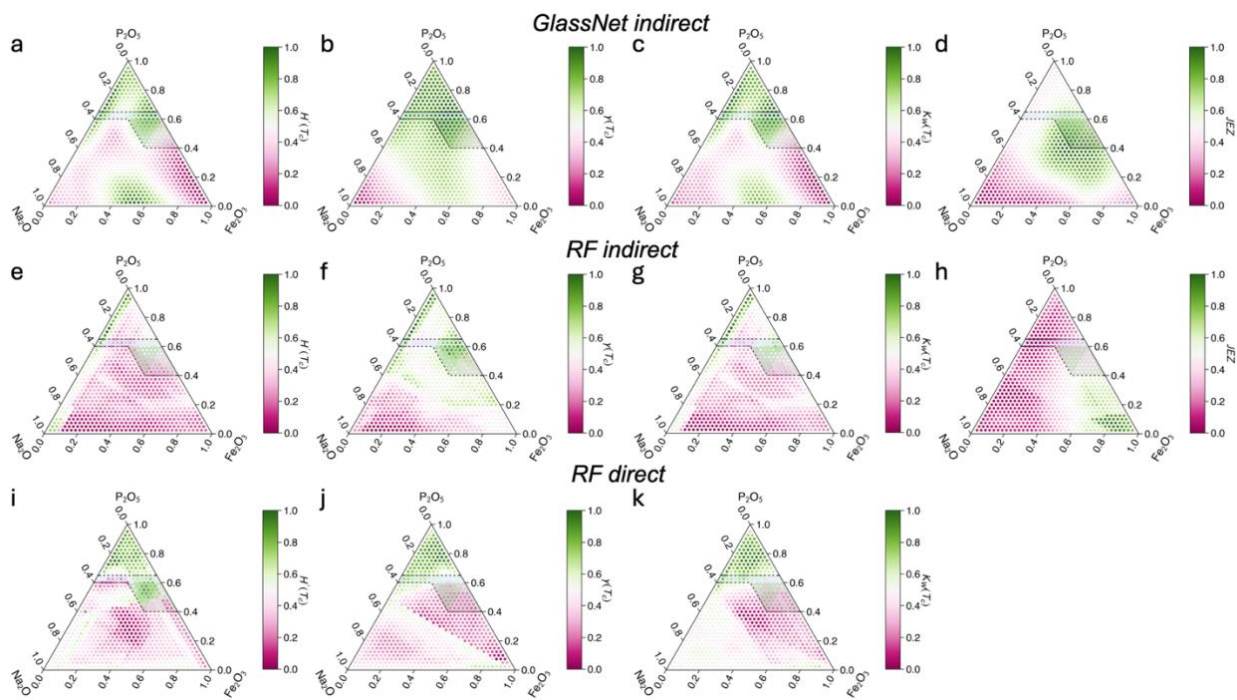

**Figure 11.** PIML predictions of GS parameters and *Jezica*, projected onto the $Na_2O$-$Fe_2O_3$-$P_2O_5$ ternary: (a-d) GlassNet (indirect) predictions, (e-h) RF indirect predictions, and (i-k) RF direct predictions. The true glass-forming region[43] is shaded in gray and bounded by a black line. Indirect predictions here refer to predictions made from a combination of the predictions for the characteristic temperatures and $\log(U_{max})$; direct predictions refer to models that directly predict a GS parameter. Note that there is no direct prediction of *Jezica* because we do not have $\eta(T_l)$ values in the training data and cannot compute "true" *Jezica* parameters for training.

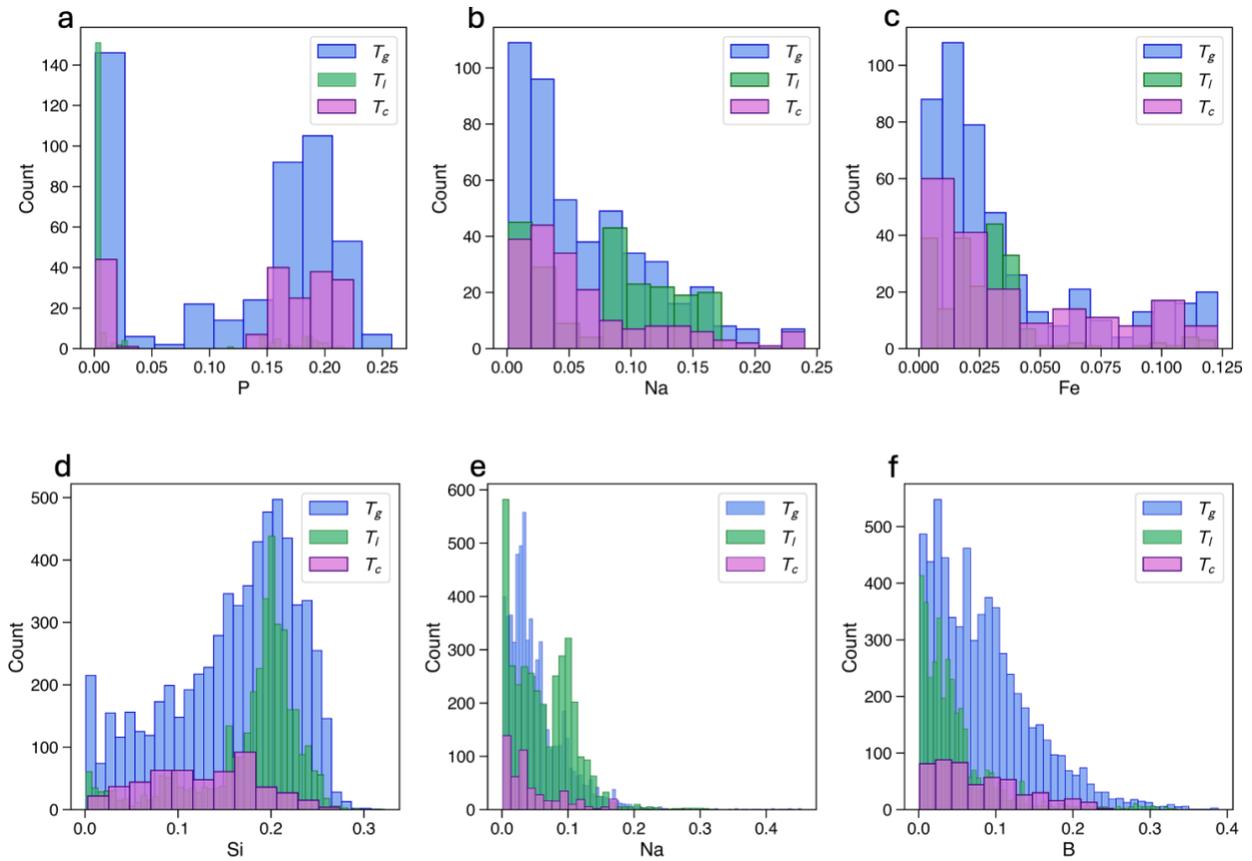

**Figure 12.** Counts of glasses containing non-zero fractions of (a) P, (b) Na, and (c) Fe in the $T_g$, $T_l$, and $T_c$ subsets and of (d) Si, (e) Na, and (f) B in the $T_g$, $T_l$, and $T_c$ subsets. These plots demonstrate the compositions and number of (a-c) $Na_2O$-$Fe_2O_3$-$P_2O_5$ glasses and (d-f) $Na_2O$-$B_2O_3$-$SiO_2$ glasses in the GlassNet data.

## Conclusion

Due to the structural complexity and vast composition space of glasses, as well as the difficulties in obtaining high quality experimental and simulation data, ML-driven optimization of glass processing and stability is still in its infancy. Many of these challenges arise from the lack of data that is representative of the true glass composition space, as the SciGlass database is currently dominated by silicates of certain compositions, while others arise from a lack of understanding of the underlying physical mechanisms of glass formation and glass stability. The data challenges are exaggerated in the type of PIML investigated here: while the ML predictions of the physical model parameters (*e.g.,* characteristic temperatures) are reasonable, the outputs of the physical models using these ML-predicted parameters are significantly less accurate than the parameter predictions. We attribute this drop in accuracy to the fact that most glasses in the SciGlass database, and consequently most of the training data used in the GlassNet and RF models assessed here, do not contain values for all parameters needed in a given physical model (*e.g.,* the all of the characteristic temperatures for GS or all of the crystal growth information needed for GFA). Thus, the distributions of the types of glasses contained in the parameter models' training data differ.

Nonetheless, the data underlying the GS parameters is much more diverse than the data underlying GFA, as the only glasses containing $T_l$, $U_{max}$, and $T_{\max(U)}$ for the GFA computation are sodium borosilicates, while there is a wide variety of glasses represented in the characteristic temperature subsets. The $H'(T_c)$ parameter is the most reliable in terms of size and diversity of underlying data and ML prediction accuracy, but should still be used with caution. We also note that the errors in the ML prediction of the top GS parameters evaluated here exhibit a correlation with the ML prediction errors for $T_c$, which could be due to the low accuracy of the $T_c$ model. We attribute the low accuracy of the $T_c$ model to two factors: *i*) the small number of glasses containing $T_c$ and *ii*) that its value can vary much more than the other characteristic temperatures for a given composition in SciGlass. While the latter (the experimental or simulation uncertainty) is much more difficult to control, the former offers a possible path forward for improving GS ML predictions: improve the $T_c$ ML models simply by collecting more $T_c$.

However, to reach quantitative accuracy of GS ML predictions, we would need all characteristic temperature values for a given glass across many glass systems, which is tedious and time-consuming, but also a process that needs to be guided in order to efficiently search the vast glass composition space. Because the results here demonstrate that random forests, from which robust uncertainty estimates are much more easily attained than neural networks, are sufficient for the prediction of GS- and GFA-relevant properties, we suggest that an uncertainty-based surrogate modeling approach may be useful in guiding the data generation and collection process. As exemplified by the ternary glass systems investigated here, a comprehensive analysis of which GS or liquid-based (*e.g.*, *Jezica*) parameter can best identify the glass-forming regions of several glass systems would not only be a useful process for generating diverse ML training data via *surrogate modeling within each glass system,* but also for *understanding the differences in glass formation mechanism of different glass families*. Furthermore, the investigation of other glass properties, such as cation field strength and optical basicity,[44] or the development of novel features such as spectral descriptors,[45] may shed light on the GFA of different glasses and may be focus of future work. Lastly, this data would need to be published according to FAIR guidelines, as opposed to only residing in tables and figures in the scientific literature, which are difficult to access and integrate into existing databases.

## Code & Data Availability

All code and related data for reproducing the results in this paper is available at https://github.com/CitrineInformatics-ERD-public/piml_glass_forming_ability.

## Acknowledgements

The information, data, or work presented herein was funded in part by the Advanced Research Projects Agency-Energy (ARPA-E), U.S. Department of Energy, under Award Number DE-AR0001613. This research used resources of the National Energy Research Scientific Computing Center, a DOE Office of Science User Facility supported by the Office of Science of the U.S. Department of Energy under Contract No. DE-AC02-05CH11231 using NERSC award NERSC DDR-ERCAP0029533. The views and opinions of authors expressed herein do not necessarily state or reflect those of the United States Government or any agency thereof. Pacific Northwest National Laboratory (PNNL) is operated by Battelle Memorial Institute for the DOE under contract DE-AC05-76RL01830. DRC acknowledges the funding of CNPq - INCT (National Institute of Science and Technology on Materials Informatics, grant n. 371610/2023-0). The authors also acknowledge Pavel Ferkl from PNNL for reviewing the manuscript and providing valuable feedback on this work.## References

1. Mauro JC, Philip CS, Vaughn DJ, Pambianchi MS. Glass science in the United States: Current status and future directions. Int J Appl Glass Sci. 2014;5(1):2–15. https://doi.org/10.1111/ijag.12058

2. Mauro JC, Zanotto ED. Two Centuries of Glass Research: Historical Trends, Current Status, and Grand Challenges for the Future. Int J Appl Glass Sci. 2014;5(3):313–327. https://doi.org/10.1111/ijag.12087

3. Jiusti J, Cassar DR, Zanotto ED. Which glass stability parameters can assess the glass-forming ability of oxide systems? Int J Appl Glass Sci. 2020;11(4):612–621. https://doi.org/10.1111/ijag.15416

4. Asayama E, Takebe H, Morinaga K. Critical cooling rates for the formation of glass for silicate melts. ISIJ Int. 1993;33(1):233–238. https://doi.org/10.2355/isijinternational.33.233

5. Fang CY, Yinnon H, Uhlmann DR. A kinetic treatment of glass formation. VIII: Critical cooling rates for $Na_2O\ SiO_2$ and $K_2O\ SiO_2$ glasses. J Non Cryst Solids. 1983;57(3):465-471. https://doi.org/10.1016/0022-3093(83)90433-7

# Supporting Information

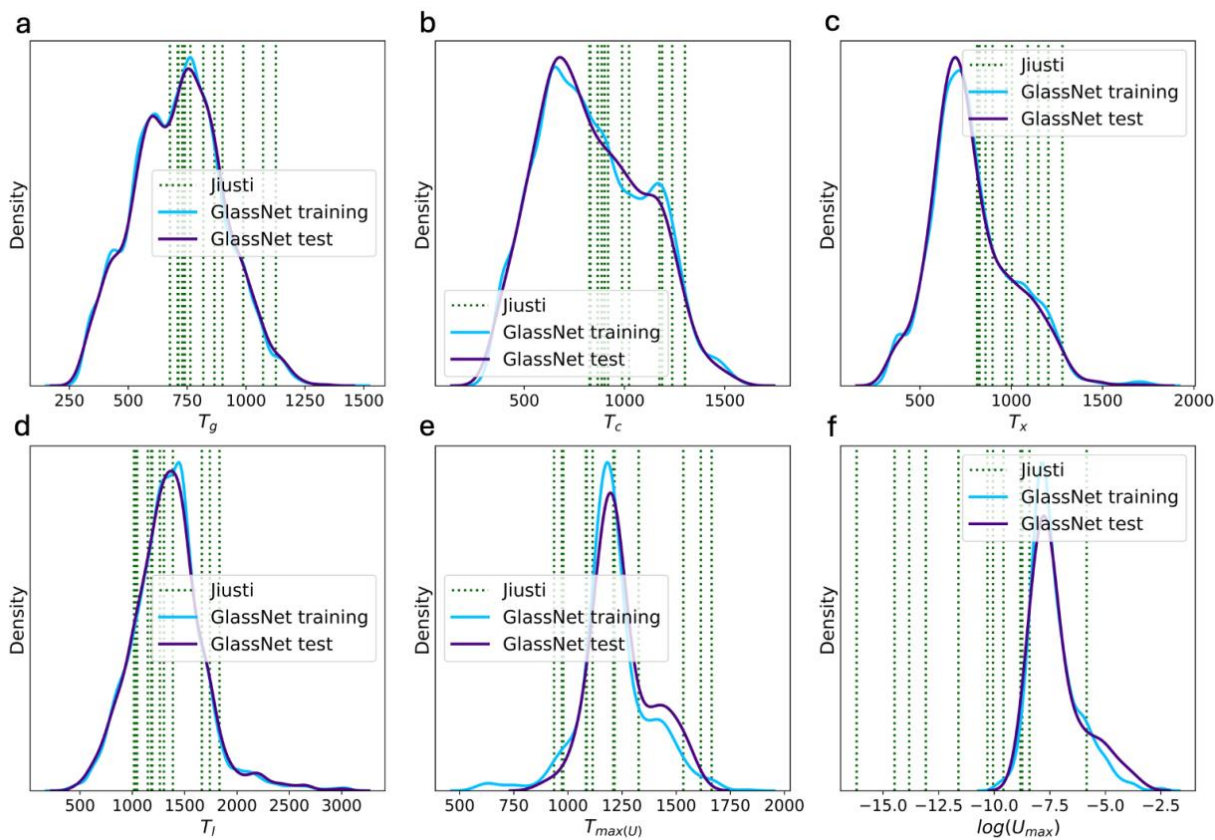

**Figure S1.** Property distributions of GlassNet training and test sets (light blue and dark purple, respectively), and property values in Jiusti dataset (dotted green lines) for properties needed to compute (a-d) GS and (e-f) GFA. The Jiusti property values are within the GlassNet distributions for all properties except $\log(U_{max})$. All temperatures are in Kelvin and $U_{max}$ is in m/s.

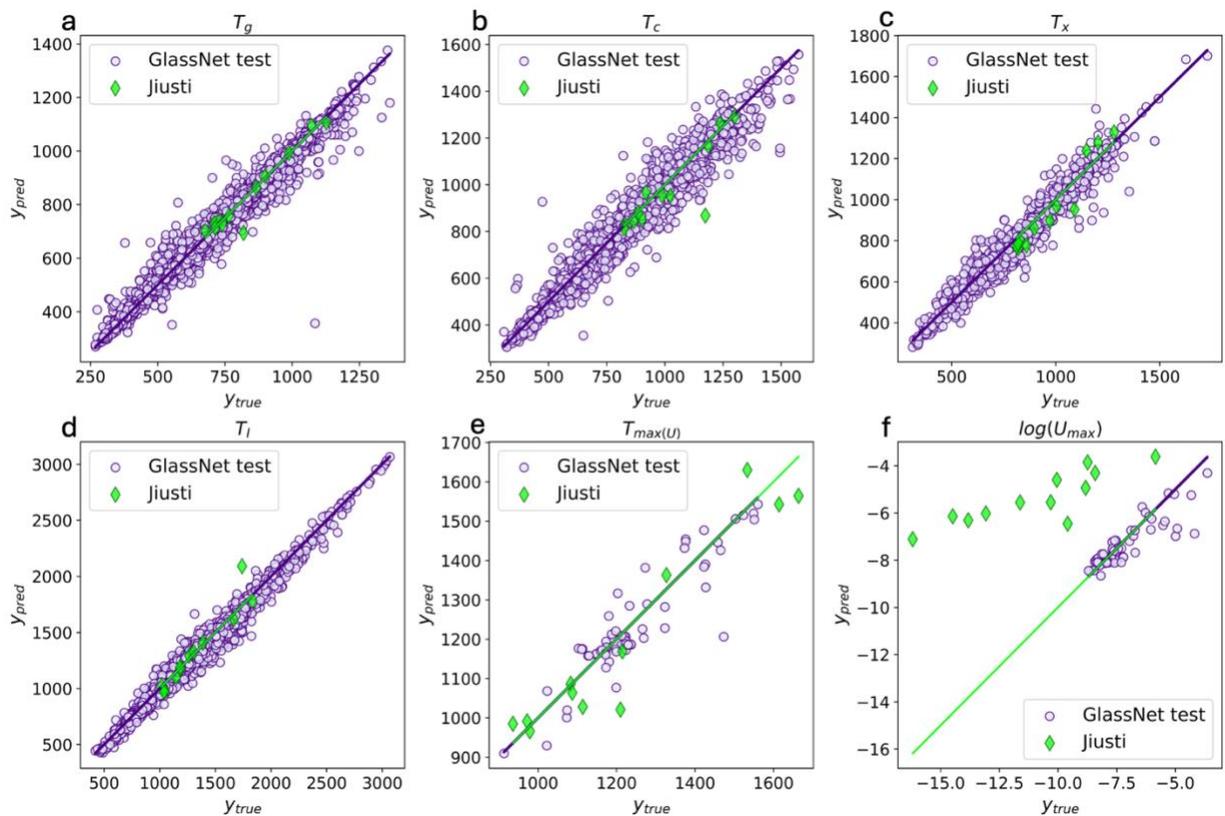

**Figure S2.** Parity plots ($y_{pred}$ vs $y_{true}$) for the properties needed to compute (a-d) GS and (e-f) GFA, as predicted by GlassNet on the GlassNet test set and the Jiusti dataset (lavender circles and green diamonds, respectively). The properties of the Jiusti data are predicted within GlassNet test set accuracy for all properties except $\log(U_{max})$, due to the distribution shift of this data relative to the GlassNet data (**Figure S1**). All temperatures are in Kelvin and $U_{max}$ is in m/s.

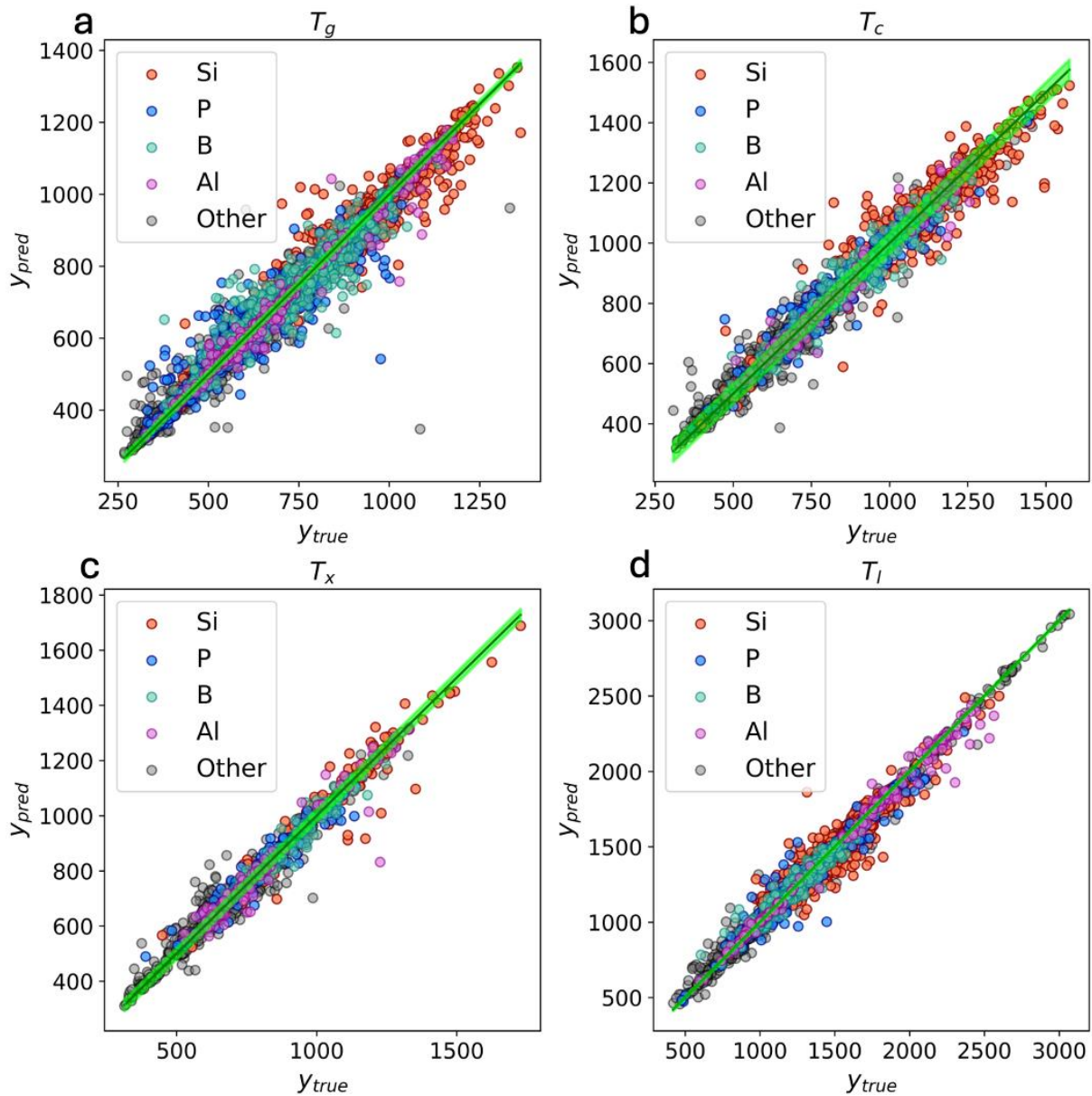

**Figure S3**. Parity plots ($y_{pred}$ vs $y_{true}$) for the GS-relevant properties (characteristic temperatures) broken down by family, as predicted by RF. Green shaded region denotes the standard deviation in these temperatures for a given composition in the SciGlass database, where we have taken the 75$^{th}$ percentile of the standard deviations over all duplicated compositions. All temperatures are in Kelvin.

**Table S1.** Number of training data points, $N_{T_i}$, and coefficients of determination, $R^2_{T_i}$, for each characteristic temperature, broken down by glass family, as predicted by RF. The last row contains the corresponding values for all glasses for reference.

|    | $N_{T_g}$ | $R^2_{T_g}$ | $N_{T_c}$ | $R^2_{T_c}$ | $N_{T_x}$ | $R^2_{T_x}$ | $N_{T_l}$ | $R^2_{T_l}$ |
|---|---|---|---|---|---|---|---|---|
| **Si** | 19,118 | 0.95 | 4,471 | 0.88 | 1,364 | 0.91 | 19,238 | 0.93 |
| **P**  | 12,074 | 0.92 | 2,538 | 0.92 | 1,402 | 0.93 | 3,117 | 0.95 |
| **B** | 14,228 | 0.93 | 2,235 | 0.90 | 1,100 | 0.94 | 4,449 | 0.96 |
| **Al** | 3,036 | 0.98 | 1,274 | 0.97 | 1,174 | 0.95 | 1,300 | 0.99 |
| **Other** | 17,638 | 0.97 | 7,564 | 0.95 | 5,477 | 0.95 | 8,108 | 0.99 |
| *All* | 66,094 | 0.97 | 18,082 | 0.96 | 10,517 | 0.96 | 36,212 | 0.97 |

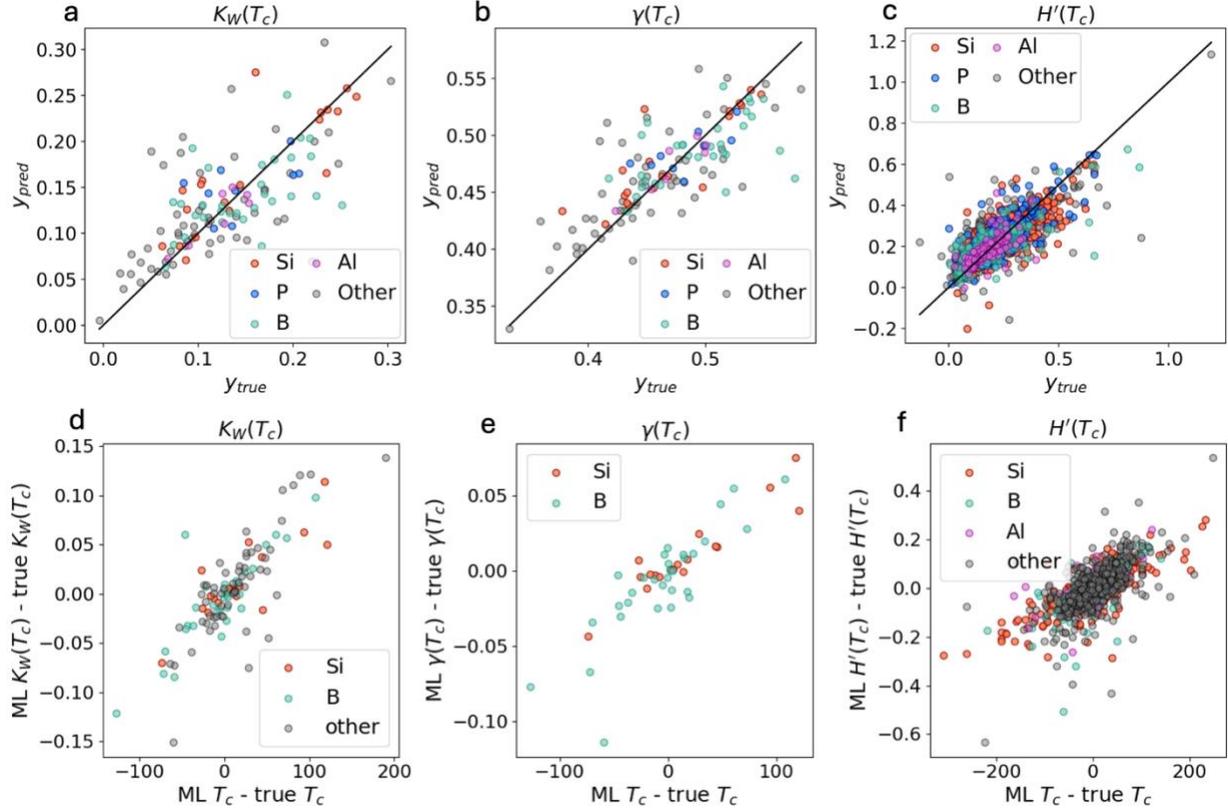

**Figure S4**. (a-c) Parity plots ($y_{pred}$ vs $y_{true}$) for the top three ML-predicted GS parameters broken down by family, as predicted by RF. (d-f) GS residuals vs. $T_c$ residuals for (d) $K_w(T_c)$, (e) $\gamma(T_c)$, and (f) $H'(T_c)$, demonstrating the correlation between the two.

**Table S2.** Coefficients of determination, $R_i^2$, and mean absolute errors, $MAE_i$, for each GS parameter, broken down by glass family. Because $R^2$ values are not well-defined for small dataset sizes, no $R^2$ are given for GS parameters with a test set size of less than ten. The last row contains the corresponding values for all glasses for reference.

| | $R^2_{K_w}$ | $MAE_{K_w}$ | $R^2_{\gamma}$ | $MAE_{\gamma}$ | $R^2_{H'}$ | $MAE_{H'}$ |
|---|---|---|---|---|---|---|
| **Si** | 0.68 | 0.026 | 0.69 | 0.017 | 0.53 | 0.060 |
| **P** | NA | 0.031 | NA | 0.017 | 0.60 | 0.050 |
| **B** | 0.20 | 0.033 | 0.26 | 0.025 | 0.61 | 0.053 |
| **Al** | NA | 0.012 | NA | 0.009 | 0.40 | 0.035 |
| **Other** | 0.36 | 0.037 | 0.56 | 0.025 | 0.59 | 0.051 |
| *All* | 0.45 | 0.035 | 0.59 | 0.024 | 0.60 | 0.052 |

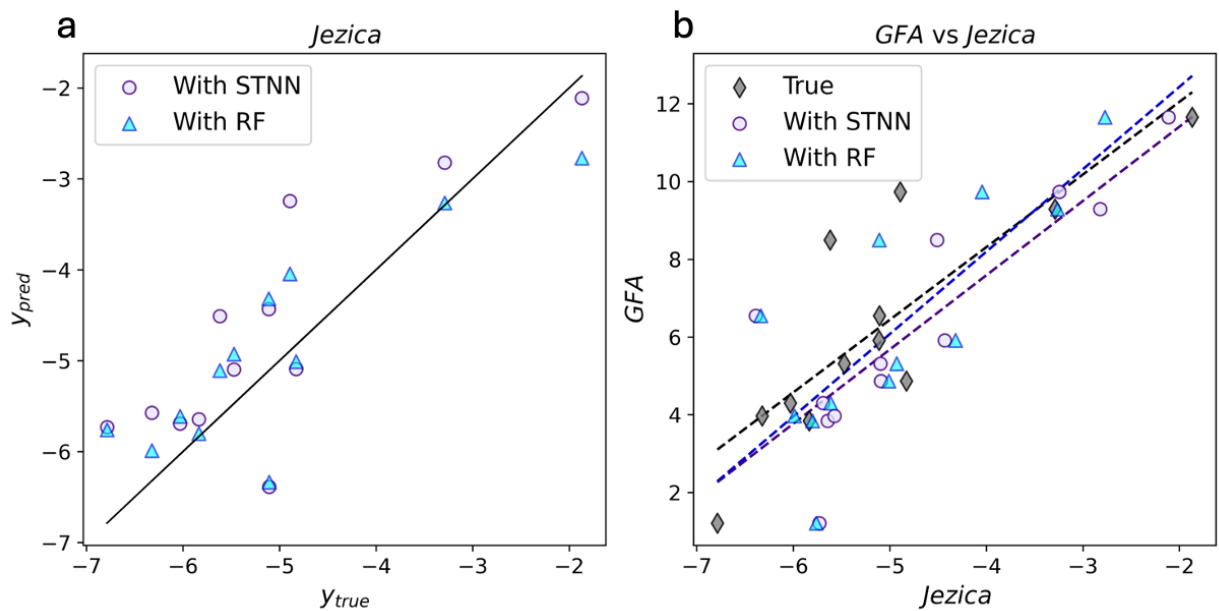

**Figure S5.** (a) Parity plot ($y_{pred}$ vs $y_{true}$) for PIML *Jezica* predictions from GlassNet and RF for the Jiusti dataset. Predictions by STNN GlassNet and RF are represented by lavender circles and cyan triangles, respectively. (b) True GFA vs. *Jezica* for the Jiusti dataset, demonstrating the correlation between the PIML *Jezica* predictions and the true GFA. True *Jezica* values are represented by gray diamonds and PIML *Jezica* predictions from STNN GlassNet and RF are represented by lavender circles and cyan triangles, respectively. The dashed lines represent lines of best fit, with all having a slope of ~2.

**Direct random forest models**

The direct random forest (RF) models used to predict GS parameters directly (*i.e.*, no physics-informed ML) to produce the bottom panels of **Figures 10-11** were built with the default parameters of scikit-learn. In these models, instead of training models and predicting each characteristic temperature, we took the subset of GlassNet data containing all the temperatures needed for a given parameter and used those to compute true values for that parameter, using this as the training data to predict the GS parameter directly. For example, for $H'(T_c) = \frac{T_c - T_g}{T_g}$, we took all glasses containing values for both $T_c$ and $T_g$, computed $H'(T_c)$, and used this data to train a single model to predict $H'(T_c)$.